\documentclass[aps,pra,twocolumn,superscriptaddress]{revtex4-1}

\usepackage{graphicx}
\usepackage{amsmath}

\begin{document}

\title{Second-order Fermionic Interference with Independent Photons}

\author{Jianbin Liu}
\email[]{liujianbin@xjtu.edu.cn}
\affiliation{Electronic Materials Research Laboratory, Key Laboratory of the Ministry of Education \& International Center for Dielectric Research, Xi'an Jiaotong University, Xi'an 710049, China}

\author{Hui Chen}
\affiliation{Electronic Materials Research Laboratory, Key Laboratory of the Ministry of Education \& International Center for Dielectric Research, Xi'an Jiaotong University, Xi'an 710049, China}

\author{Yu Zhou}
\affiliation{MOE Key Laboratory for Nonequilibrium Synthesis and Modulation of Condensed Matter, Department of Applied Physics, Xi'an Jiaotong University, Xi'an 710049, China}

\author{Huaibin Zheng}
\affiliation{Electronic Materials Research Laboratory, Key Laboratory of the Ministry of Education \& International Center for Dielectric Research, Xi'an Jiaotong University, Xi'an 710049, China}

\author{Fu-li Li}
\affiliation{MOE Key Laboratory for Nonequilibrium Synthesis and Modulation of Condensed Matter, Department of Applied Physics, Xi'an Jiaotong University, Xi'an 710049, China}

\author{Zhuo Xu}
\affiliation{Electronic Materials Research Laboratory, Key Laboratory of the Ministry of Education \& International Center for Dielectric Research, Xi'an Jiaotong University, Xi'an 710049, China}

\date{\today}

\begin{abstract}
The experimental study of the second-order interference with fermions is much less than the one with bosons since it is much more difficult to do experiments with fermions than with photons. Based on the conclusion that the behavior of two identical classical particles has exactly half fermionic and half bosonic characteristics (PRA \textbf{88}, 012130 (2013)), we have studied the second-order interference of fermions via the second-order interference of photons in Hanbury Brown-Twiss and Hong-Ou-Mandel interferometers, respectively. The experimental results are consistent with the theoretical predictions, which serve as an efficient guidance for the future interference experiments with fermions. The employed method offers an interesting and convenient way to study the coherence of fermions.
\end{abstract}

\maketitle

\section{Introduction}\label{introduction}

Particles can be categorized into two types based on the spin they have \cite{note-anyons}. One is boson, which has zero or integral spin and follows Bose-Einstein statistics. The other one is fermion, which has half-integral spin and follows Fermi-Dirac statistics \cite{dirac}. The coherence of particles is an important topic in both quantum optics and quantum information \cite{mandel-book,nielsen}. Photon is massless boson and the coherence properties of photon have been well understood via different interference experiments. For instance, the first-order interference pattern can be formulated when there is only one photon in a interferometer at one time \cite{aspect}. Photons emitted by thermal light source are not independent as people thought before 1950s \cite{hbt}. Fast development of modern technology makes it convenient to do experiments with photons \cite{bachor}. However, similar experiments may be challenge if massive particles are employed. For example, experimental realization of laser-like state with massive bosons is long after the invention of laser \cite{bec-1,bec-2}.

If there is some relation between the coherence of massive particles and photons, the coherence properties of massive particles can be predicted based on the properties of photons. In the first-order interference, the coherence properties of massive bosons and fermions are similar as the ones of photons. For instance, the first-order interference pattern of electrons (fermion) in a Young's double-slit interferometer is similar as the one of photons \cite{tonomura}. There is first-order interference pattern by superposing two independent Bose-Einstein Condensations (BEC) \cite{andrews}, which is similar as the one of superposing two independent lasers \cite{mandel-laser}.  However, things become different for the second- and higher-order interference of massive particles. The interference patterns of massive bosons are similar as the ones of photons in the second- and higher-order interference. There is two-atom bunching in ultracold quantum gases above the threshold temperature of BEC while no bunching for atoms in BEC \cite{schellekens} in a Hanbury Brown and Twiss (HBT) interferometer, which is similar as there is two-photon bunching for photons in thermal light \cite{hbt} and no bunching for photons in single-mode laser light \cite{glauber}. Two-particle antibunching was observed in a HBT interferometer with thermal electrons \cite{henny,oliver}, which is different from two-photon bunching of thermal light.

There are limited number of second-order interference experiments with fermions \cite{henny,oliver,liu,rom,iannuzzi,jeltes,schmidt,neder,bocquillon} due to the experiments with fermions are challenge. If photons can be employed to study the second-order interference of fermions, it will become easier to understand the second-order coherence of fermions. Entangled photon pairs have been employed to simulate the second-order interference of fermions \cite{zeilinger,omar,sansoni,peruzzo,matthews,tichy,crespi}. Special attention is needed to employ entangled photon pairs to simulate the interference of fermions since only antisymmetrical state is possible \cite{zeilinger}. On the other hand, directly generating entangled photons with more than two photons is still challenge \cite{hubel}. It is tempting to simulate the second- and higher-order interference of fermions with independent photons. Recently, T\"{o}ppel \textit{et.al.} pointed out that \lq\lq the quantum behavior of a pair of identical classical particles has exactly half fermionic and half bosonic characteristics\rq\rq \cite{toppel}. Their conclusion is proved for identical particles in two-particle states in a multi input-output ports interferometer.  In this paper, we will prove that similar conclusion is true for non-identical particles in thermal state via Feynman's path integral theory, which means it is possible to employ photons in classical state to study the second-order interference of fermions.

The paper is organized as follows. In Sect. \ref{theory}, we will employ Feynman's path integral theory to reproduce the conclusion given in T\"{o}ppel \textit{et.al.}'s paper in HBT and Hong-Ou-Mandel (HOM) interferometers for non-identical particles. The second-order interference experiments with photons in thermal state to study the behavior of fermions are in Sect. \ref{experiments}. The discussions and applying our results to interpret the existed second-order interference experiments with real fermions are in Sect. \ref{discussions}. Section \ref{conclusion} summarizes our conclusions.

\section{Theory}\label{theory}

In T\"{o}ppel \textit{et.al.}'s paper \cite{toppel}, they employed second-quantization formalism to prove that for two identical particles in a multi input-output ports interferometer, the behavior of a pair of identical classical particles has exactly half fermionic and half bosonic characteristics. They also mentioned that it is possible to simulate the behavior of unentangled fermions simply with unentangled bosons. According to Pauli exclusion principle, it is impossible for two fermions to occupy exactly the same one-particle state at the same time, which means it is impossible for two identical fermions in a HBT interferometer to trigger a two-particle coincidence count. However, two-particle antibunching was observed for fermions in a HBT interferometer \cite{iannuzzi,jeltes}. The physics behind is that identical particles are not necessary for two-particle interference. The necessary and sufficient condition for two-particle interference is there exist more than one different yet indistinguishable alternatives for two particles to trigger a two-particle coincidence count \cite{feynman,feynman-qpt}. Non-identical particles can have indistinguishable alternatives. We have experimentally observed two-photon interference with photons of frequency difference equals more than 200 MHz, where the frequency bandwidth of each light beam is about 200 kHz \cite{liu-oc-2015,liu-cpb}. There is two-particle interference as long as the different alternatives to trigger a two-particle coincidence count are, in principle, indistinguishable for the employed detection system \cite{liu-oc-2015}. In the following part of this section, we will calculate the second-order coherence functions of thermal fermions in HBT and HOM interferometers, respectively.

\subsection{HBT interferometer}

\begin{figure}[htb]
\centering
\includegraphics[width=60mm]{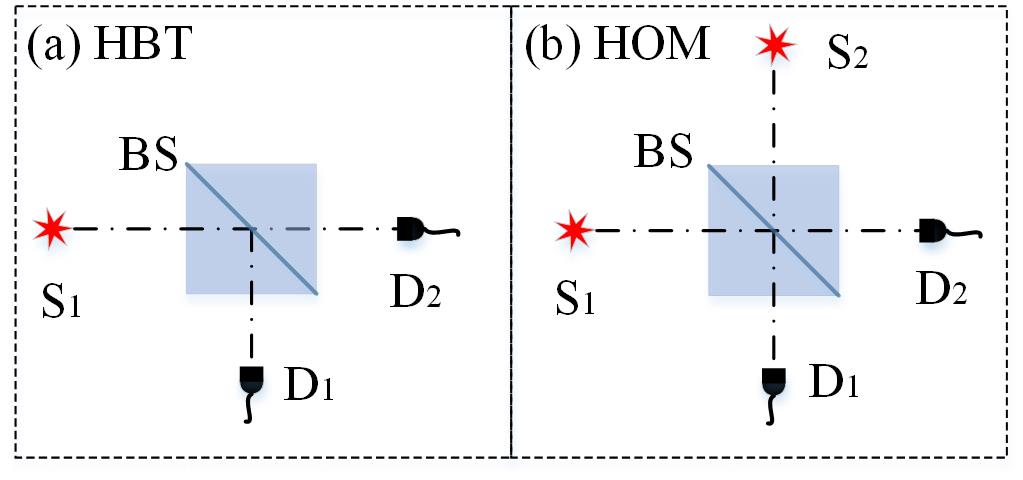}%
\caption{(Color online) HBT and HOM interferometers. S$_1$ and S$_2$ are two particle sources. BS is a 50:50 non-polarized beam splitter. D$_1$ and D$_2$ are two single-particle detectors.}\label{1}
\end{figure}

HBT interferometer is shown in Fig. \ref{1}(a), which was firstly employed by Hanbury Brown and Twiss to measure the angular size of Sirius \cite{hbt-1}. It is the simplest second-order interferometer and often employed to study the statistics of photons in different states \cite{mandel-book}. S$_1$ is a particle source. Fermions can not form coherent state as bosons and the emitted particles are assumed to be in thermal state in the following calculations. BS is a 50:50 non-polarized beam splitter. D$_1$ and D$_2$ are two single-particle detectors. The output signals of these two detectors are sent to a coincidence counting system to measure two-particle coincidence counts, which is not shown. There are two different alternatives for two particles in thermal state to trigger a coincidence count in Fig. \ref{1}(a). One is particle a goes to D$_1$ and particle b goes to D$_2$. The other one is particle b goes to D$_1$ and particle a goes to D$_2$. If these two different alternatives are indistinguishable, the second-order coherence function of thermal bosons in a HBT interferometer is \cite{feynman,feynman-qpt}
\begin{equation}\label{HBT-b-1}
G^{(2)}_B(\mathbf{r}_1,t_1;\mathbf{r}_2,t_2)=\langle |A_{a1}A_{b2}+A_{a2}A_{b1}|^2 \rangle,
\end{equation}
where $\langle...\rangle$ means ensemble average, {\it i.e.}, taking all possible realizations into account. For a stationary and ergodic process, the ensemble average is equivalent to a time average over a long period \cite{mandel-book}. $(\mathbf{r}_1,t_1)$ and $(\mathbf{r}_2,t_2)$ are the space-time coordinates for the particle detection events at D$_1$ and D$_2$, respectively. $A_{\alpha\beta}$ is the probability amplitude for particle $\alpha$ goes to Detector $\beta$ ($\alpha=a$ and $b$, $\beta=1$ and 2).

When the particles emitted by S$_1$ are thermal fermions, the second-order coherence function in a HBT interferometer is \cite{feynman,feynman-qpt}
\begin{equation}\label{HBT-f-1}
G^{(2)}_F(\mathbf{r}_1,t_1;\mathbf{r}_2,t_2)=\langle |A_{a1}A_{b2}-A_{a2}A_{b1}|^2 \rangle.
\end{equation}
The two different alternatives to trigger a two-particle coincidence count are also assumed to be indistinguishable. The meanings of the symbols in Eq. (\ref{HBT-f-1}) are similar as the ones in Eq. (\ref{HBT-b-1}). The only difference between these two equations is the plus sign in Eq. (\ref{HBT-b-1}) becomes minus sign in Eq. (\ref{HBT-f-1}), which is due to bosons and fermions obey different exchanging symmetries \cite{dirac,feynman}.

When the particles emitted by S$_1$ are classical particles, these two different alternatives are distinguishable. The second-order coherence function in Fig. \ref{1}(a) is \cite{feynman,feynman-qpt}
\begin{equation}\label{HBT-c-1}
G^{(2)}_C(\mathbf{r}_1,t_1;\mathbf{r}_2,t_2)=\langle |A_{a1}A_{b2}|^2+|A_{a2}A_{b1}|^2 \rangle,
\end{equation}
where the probabilities instead of probability amplitudes are summed to calculate the second-order coherence function in the classical particle case.

Combining Eqs. (\ref{HBT-b-1}) - (\ref{HBT-c-1}), it is straightforward to prove that
\begin{eqnarray}\label{HBT-c-2}
&&G^{(2)}_C(\mathbf{r}_1,t_1;\mathbf{r}_2,t_2)\nonumber\\
=&&\frac{1}{2}[G^{(2)}_B(\mathbf{r}_1,t_1;\mathbf{r}_2,t_2)+G^{(2)}_F(\mathbf{r}_1,t_1;\mathbf{r}_2,t_2)]
\end{eqnarray}
is satisfied for particles in thermal state in a HBT interferometer, which is similar as T\"{o}ppel \textit{et.al.}'s conclusion \cite{toppel}. With some re-arrangement, we can have the second-order coherence function of fermions in a HBT interferometer via the second-order coherence functions of bosons and classical particles, \textit{i.e.},
\begin{eqnarray}\label{HBT-f-3}
&&G^{(2)}_F(\mathbf{r}_1,t_1;\mathbf{r}_2,t_2)\nonumber\\
=&&2G^{(2)}_C(\mathbf{r}_1,t_1;\mathbf{r}_2,t_2)-G^{(2)}_B(\mathbf{r}_1,t_1;\mathbf{r}_2,t_2).
\end{eqnarray}
Equation (\ref{HBT-f-3}) can be further simplified if the type of particle is specified. In order to compare the theoretical results with our experimental results in Sect. \ref{experiments}, we will give the second-order coherence function for photons. The second-order coherence function of thermal light has been calculated in many textbooks of quantum optics, for instance, see \cite{mandel-book,shih-book}. We will not repeat the calculation process and directly employ the final results. The normalized second-order coherence function always equals to 1  for classical particles since two single-particle detection events are independent \cite{glauber}. With the help of Eq. (\ref{HBT-f-3}) and coherence functions for bosons and classical particles, the normalized second-order temporal coherence function of \lq\lq fermionic photon\rq\rq \cite{note-f} in a HBT interferometer is
\begin{eqnarray}\label{HBT-f-4}
&&g^{(2)}_F(t_1-t_2)=1-\text{sinc}^2[\pi \Delta\nu (t_1-t_2)],
\end{eqnarray}
where the positions of these two detectors are assumed to be the same in order to concentrate on temporal coherence function. $\text{sinc}(x)$ equals $\sin(x)/x$. $t_1$ and $t_2$ are the detection time of particles by D$_1$ and D$_2$, respectively. $\Delta\nu$ is the frequency bandwidth of \lq\lq fermionic photon\rq\rq, which is the same as the frequency bandwidth of the employed photons.

With the same method above, the normalized one-dimensional second-order spatial coherence function of \lq\lq fermionic photon\rq\rq in a HBT interferometer is \cite{shih-book}
\begin{eqnarray}\label{HBT-f-5}
&&g^{(2)}_F(x_1-x_2)=1-\text{sinc}^2[\frac{\pi l}{z \lambda}(x_1-x_2)],
\end{eqnarray}
where paraxial approximation have been assumed to simplify the expression. $x_1$ and $x_2$ are the transverse coordinates of D$_1$ and D$_2$ in the detection planes, respectively. $l$ is the size of thermal light source. The distance between S$_1$ and D$_1$ planes equals the one between S$_1$ and D$_2$ planes, which is written as $z$. $\lambda$ is the wavelength of photon.

\subsection{HOM interferometer}

Figure \ref{1}(b) is a Shih-Alley or HOM interferometer \cite{shih,hom}, which is an important tool to measure the indistinguishability of photons \cite{santori} and a basic element in quantum information \cite{nielsen}. We will prove that similar conclusion as T\"{o}ppel \textit{et.al.}'s holds for the second-order interference of two independent thermal particle beams in a HOM interferometer.

There are eight different ways to trigger a two-particle coincidence count for two independent thermal beams in Fig. \ref{1}(b) \cite{liu-oe-2013}, which are A$_{1a1}$A$_{1b2}$, A$_{1a2}$A$_{1b1}$, A$_{2a1}$A$_{2b2}$, A$_{2a2}$A$_{2b1}$, A$_{1a1}$A$_{2b2}$, A$_{1a2}$A$_{2b1}$, A$_{2a1}$A$_{1b2}$, and A$_{2a2}$A$_{1b1}$, respectively. The meaning of A$_{\alpha\beta\gamma}$ is the probability amplitude that particle $\beta$ emitted by source $\alpha$ goes to detector $\gamma$ ($\alpha=1$ and $2$, $\beta=a$ and $b$, and $\gamma=1$ and $2$.) For example, there are two different ways to trigger a two-particle coincidence count when these two particles are emitted by S$_1$, A$_{1a1}$A$_{1b2}$ and A$_{1a2}$A$_{1b1}$. A$_{1a1}$A$_{1b2}$ is the probability amplitude that particle $a$ goes to D$_1$ and particle $b$ goes to D$_2$. A$_{1a2}$A$_{1b1}$ corresponds to the probability amplitude that particle $a$ goes to D$_2$ and particle $b$ goes to D$_1$. Other symbols have similar meanings as the ones above.

When the particles emitted by S$_1$ and S$_2$ are thermal bosons and these eight different alternatives are indistinguishable, the second-order coherence function is \cite{liu-oe-2013,feynman-qpt}
\begin{eqnarray}\label{HOM-B-1}
&&G^{(2)}_B(\mathbf{r}_1,t_1;\mathbf{r}_2,t_2)\nonumber\\
=&&\langle |\sum_{m,n=1,2}({A_{ma1}A_{nb2}+A_{ma2}A_{nb1}})|^2 \rangle.
\end{eqnarray}
Assuming the initial phases of the particles in thermal state are random \cite{loudon}, all the cross terms in Eq. (\ref{HOM-B-1}) with different values of $m$ and $n$ will disappear after ensemble average \cite{liu-oe-2013}. Equation (\ref{HOM-B-1}) can be simplified as
\begin{eqnarray}\label{HOM-B-2}
&&G^{(2)}_B(\mathbf{r}_1,t_1;\mathbf{r}_2,t_2)\nonumber\\
=&& \sum_{m,n=1,2}\langle |{A_{ma1}A_{nb2}+A_{ma2}A_{nb1}}|^2 \rangle.
\end{eqnarray}

With the same method above, the second-order coherence function of two independent thermal fermion beams in a HOM interferometer is
\begin{eqnarray}\label{HOM-F-1}
&&G^{(2)}_F(\mathbf{r}_1,t_1;\mathbf{r}_2,t_2)\nonumber\\
=&&\langle |\sum_{m,n=1,2}({A_{ma1}A_{nb2}-A_{ma2}A_{nb1}})|^2 \rangle,
\end{eqnarray}
where these eight different alternatives are also assumed to be indistinguishable. The plus sign changes into minus sign when only the order of detectors is exchanged. Other terms are summed together \cite{feynman-qpt}. Assuming the initial phases of the fermions in thermal state are also random, Eq. (\ref{HOM-F-1}) can be simplified as
\begin{eqnarray}\label{HOM-F-2}
&&G^{(2)}_F(\mathbf{r}_1,t_1;\mathbf{r}_2,t_2)\nonumber\\
=&& \sum_{m,n=1,2}\langle|{A_{ma1}A_{nb2}-A_{ma2}A_{nb1}}|^2 \rangle.
\end{eqnarray}

When the particles emitted by the sources in Fig. \ref{1}(b) are classical particles, all the eight different alternatives are distinguishable. The second-order coherence function for classical particles is
\begin{eqnarray}\label{HOM-C-1}
&&G^{(2)}_C(\mathbf{r}_1,t_1;\mathbf{r}_2,t_2)\nonumber\\
=&& \langle \sum_{m,n=1,2}(|{A_{ma1}A_{nb2}|^2+|A_{ma2}A_{nb1}}|^2 )\rangle.
\end{eqnarray}

The same relation as T\"{o}ppel \textit{et.al.}'s conclusion \cite{toppel} holds for the second-order coherence functions expressed in Eqs. (\ref{HOM-B-2}), (\ref{HOM-F-2}), and (\ref{HOM-C-1}). The second-order coherence function of fermions in a HOM interferometer can be expressed by the coherence functions of bosons and classical particles via Eq. (\ref{HBT-f-3}). It worth noting that we only proved this equation for particles in thermal state. Whether the same relation holds for particles in other states in a HOM interferometer can be testified with the same method above.

In order to compare the theoretical and experimental results, we will give the second-order coherence function for \lq\lq fermionic photon\rq\rq in a HOM interferometer. Assuming the distances between the source and detection planes are all equal to $z$.  With the second-order coherence function of two independent thermal light beams in a HOM interferometer calculated in Ref. \cite{liu-oe-2013} and the normalized second-order coherence function of classical particles equals 1, the normalized second-order spatial coherence function of two thermal fermion beams in a HOM interferometer is
\begin{eqnarray}\label{HOM-F-3}
&&g^{(2)}_F(x_1-x_2)\nonumber\\
=&&1-\frac{1}{2}\text{sinc}^2[\frac{\pi l}{\lambda z }(x_1-x_2)]\nonumber\\
&+&\frac{1}{2}\text{sinc}^2[\frac{\pi l}{\lambda z}(x_1-x_2)]\cos[\frac{2\pi d}{\lambda z}(x_1-x_2)],
\end{eqnarray}
where one-dimension and paraxial approximation have been assumed. $x_1$ and $x_2$ are the transverse coordinates of D$_1$ and D$_2$ in the detection planes, respectively. $l$ is the size of both sources. $d$ is the distance between the middle points of S$_1$ and the image of S$_2$ vis BS$_2$.

\section{Experiments}\label{experiments}

In Sect. \ref{theory}, we have proved that the second-order coherence function of fermions in a HBT (HOM) interferometer can be expressed by the second-order coherence functions of bosons and classical particles. In this section, we will employ pseudothermal light to experimentally study the second-order interference of fermions in these two interferometers. Pesudothermal light is usually generated by impinging single-mode continuous-wave laser light onto a rotating ground glass \cite{martienssen}. The reasons why we choose pseudothermal light instead of true thermal light are as follows. The coherence properties of pseudothermal light are the same as the ones of true thermal light except the coherence time and degeneracy factor of psudothermal light can be controlled easily, which makes it suitable to experimentally study the second-order interference of photons. For instance, most ghost imaging experiments with thermal light are done with pseudothermal light \cite{shih-book,shapiro}. When photons are in the same coherence volume, all the photons are indistinguishable \cite{mandel-book}, which can be treated as bosons. Photons are distinguishable when they are in different coherence volumes \cite{mandel-book}, which can be treated as classical particles. Hence the coherence functions of bosons and classical particles can be measured in a single experiment.

\begin{figure}[htb]
\centering
\includegraphics[width=70mm]{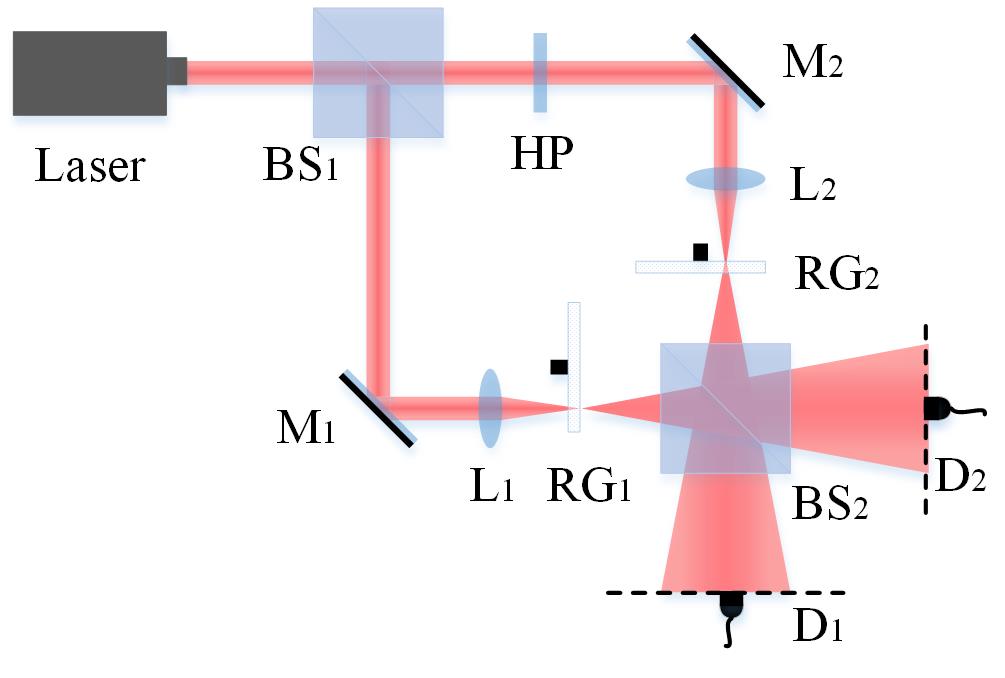}%
\caption{(Color online) Experimental setup to study the second-order interference of fermions with photons. Laser: 780 nm single-mode continuous-wave laser with frequency bandwidth 200 kHz. BS: 50:50 nonpolarized beam splitter. HP: half wave plate. M: mirror. L: Lens. RG: rotating ground glass. D: single-particle detector.}\label{2}
\end{figure}

The experimental setup to study the second-order interference of fermions with photons is shown in Fig. \ref{2}, which is the same as the one in our earlier study \cite{liu-oe-2013}. The laser employed is a single-mode continuous-wave laser with central wavelength at 780 nm and frequency bandwidth of 200 kHz (Newport, SWL-7513). BS$_1$ and BS$_2$ are two 50:50 nonpolarized beam splitters. HP is a half wave plate to change the polarization of light passing through it. The laser light is focused by lens (L) and impinged onto a rotating ground glass (RG) to generate pseudothermal light. The spatial and temporal coherence of pseudothermal light can be changed by varying the size of light spot on RG and the rotating speed of RG, respectively \cite{martienssen}. The degeneracy factor can be changed by tuning the intensity of the input laser light \cite{mandel-book}. D$_1$ and D$_2$ are two single-photon detectors (PerkinElmer, SPCM-AQRH-14-FC). The output signals of these two detectors are sent to a coincidence counting system (Becker \& Hickl GmbH, SPC630) to measure the coincidence counts, which is not shown in the figure. The distances between the source and detection planes are all equal to 910 mm.

\begin{figure}[htb]
\centering
\includegraphics[width=70mm]{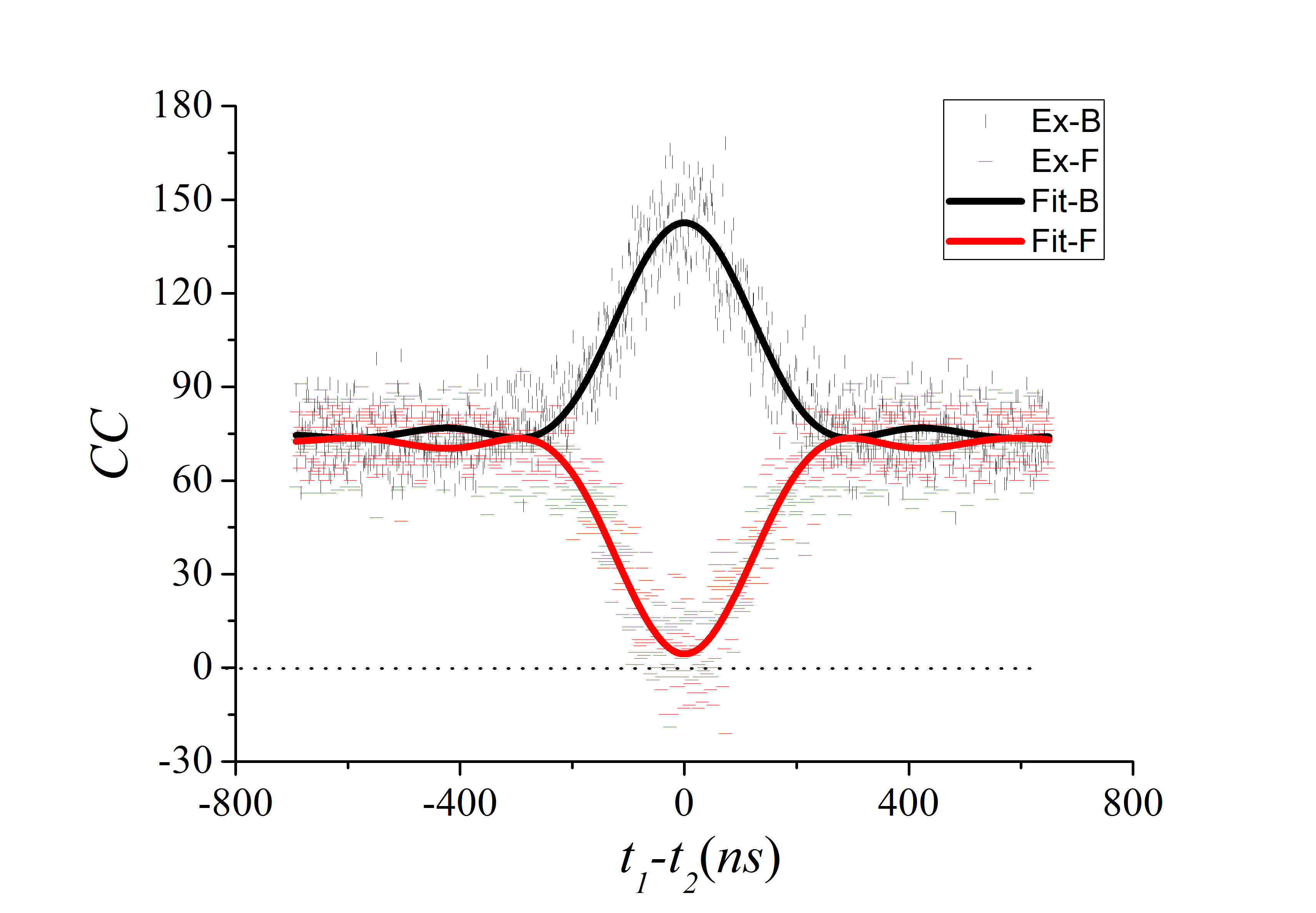}%
\caption{(Color online) Measured coincidence counts for photons (vertical black short lines) and calculated coincidence counts for fermions (horizontal red short lines).  $CC$: coincidence counts. $t_1-t_2$: time difference between two single-photon detection events in a two-photon coincidence count. Ex-B: Experimental results for bosons. Ex-F: Experimental results for fermions. Fit-B: Theoretical fit for bosons. Fit-F: Theoretical fit for fermions. }\label{3}
\end{figure}

We first measure the second-order temporal coherence function of pseudothermal light by blocking the beam reflected by M$_2$. The measured temporal coincidence counts are shown in Fig. \ref{3}. $CC$ is two-particle coincidence counts. $t_1-t_2$ is the time difference between two single-photon detection events in a two-photon coincidence count. The horizontal red short lines are coincidence counts of \lq\lq fermionic photon\rq\rq in thermal state, which are calculated by employing Eq. (\ref{HBT-f-3}). $G^{(2)}_C(t_1-t_2)$ is a constant and proportional to the background coincidence counts. $G^{(2)}_B(t_1-t_2)$ is proportional to the measured coincidence counts \cite{glauber}. The black curve is the fitting of measured coincidence counts by employing the second-order temporal coherence function of thermal light \cite{shih-book}. The red curve is the fitting of calculated coincidence counts for fermions by employing Eq. (\ref{HBT-f-4}). The fitted second-order coherence time of both curves are 296 ns. The data below the dot line in Fig. \ref{3} should be ignored, for the measured coincidence counts can not be less than 0 in the experiments with real fermions. The calculated coincidence counts for fermions is similar as the measured two-particle coincidence counts of thermal neutrons in a HBT interferometer \cite{iannuzzi}.

\begin{figure}[htb]
\centering
\includegraphics[width=70mm]{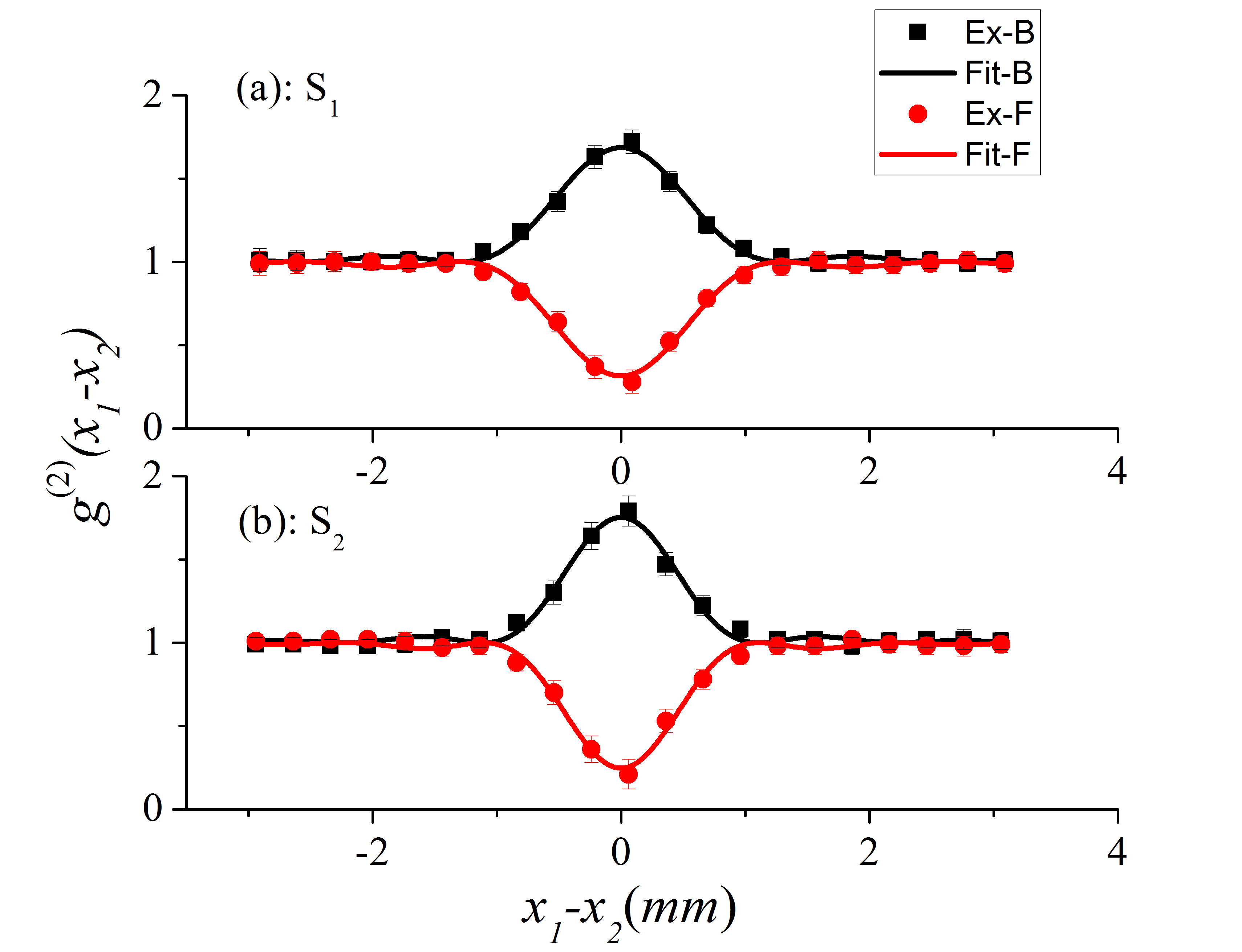}%
\caption{(Color online) Measured spatial coherence functions for photons (black squares) and calculated spatial coherence functions for fermions (red circles) in a HBT interferometer. (a) and (b) are the measured results for source 1 and source 2, respectively. $g^{(2)}(x_1-x_2)$: normalized second-order coherence function. $x_1-x_2$: transverse coordinate difference between two single-photon detectors. Ex-B: Experimental results for bosons. Ex-F: Experimental results for fermions. Fit-B: Theoretical fit for bosons. Fit-F: Theoretical fit for fermions. }\label{4}
\end{figure}

Figures \ref{4}(a) and (b) are the measured second-order spatial coherence functions of particles emitted by S$_1$ and S$_2$, respectively. $g^{(2)}(x_1-x_2)$ is the normalized second-order coherence function and $x_1-x_2$ is the transverse coordinate difference between two detectors. The black squares are measured results for photons and the red circles are calculated results for fermions. The two-particle coincidence time window is 61 ns, which is less than the coherence time of thermal state \cite{note}. The coherence functions in Figs. \ref{4}(a) and (b) are measured by blocking the light beam reflected by M$_2$ and M$_1$, respectively. The spatial function is measured by transversely scanning the position of D$_1$ while fixing the position of D$_2$. The normalized second-order coherence function is calculated by employing the definition \cite{glauber}
\begin{equation}\label{definition}
g^{(2)}(\mathbf{r}_1,t_1;\mathbf{r}_2,t_2)=\frac{G^{(2)}(\mathbf{r}_1,t_1;\mathbf{r}_2,t_2)}{G^{(1)}(\mathbf{r}_1,t_1)G^{(1)}(\mathbf{r}_2,t_2)},
\end{equation}
where $G^{(1)}(\mathbf{r}_j,t_j)$ is the measured single-photon counting rate of D$_j$ at $(\mathbf{r}_j,t_j)$ ($j=1$ and 2). The black and red curves are theoretical fittings by employing spatial coherence functions of thermal light and Eq. (\ref{HBT-f-5}), respectively. The fitted sizes of S$_1$ and S$_2$ are 0.55 mm and 0.64 mm, respectively. The spatial coherence function of fermions in Fig. \ref{3} is the same as the one of $^3$He in thermal state measured in Ref. \cite{jeltes}. The visibility of the measured dips in Fig. \ref{4}(a) and (b) is 52.14\% and 60.13\%, respectively, which is much higher than the one in Ref. \cite{jeltes}.

\begin{figure}[htb]
\centering
\includegraphics[width=70mm]{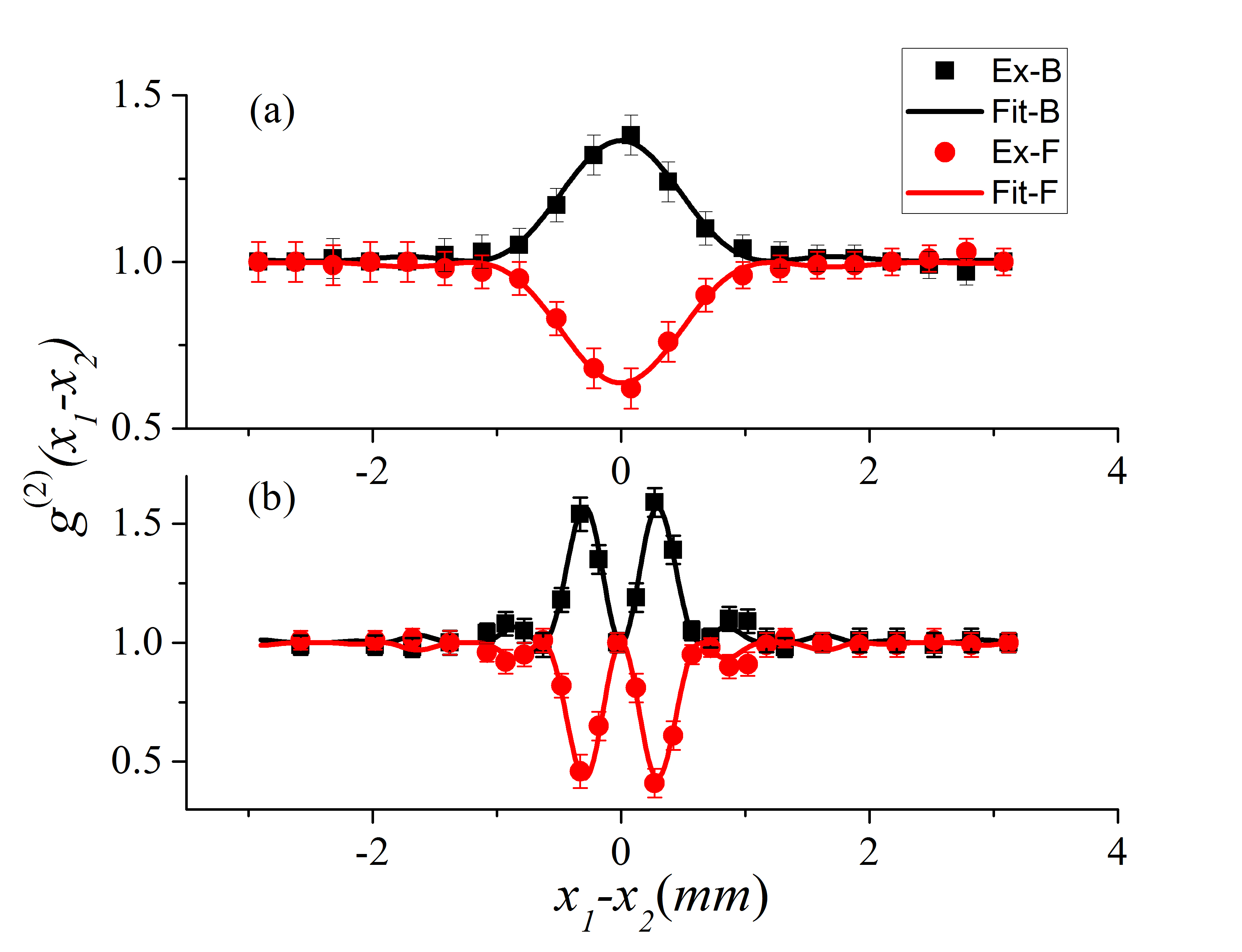}%
\caption{(Color online) Measured spatial coherence functions for photons (black squares) and calculated spatial coherence functions for fermions (red circles) in a HOM interferometer. (a) and (b) are experimental results for photons with orthogonal and parallel polarizations, respectively. The meaning of the symbols are the same as ones in Fig. \ref{4}. }\label{5}
\end{figure}

Figure \ref{5} shows the measured second-order interference patterns of two thermal beams in a HOM interferometer. The polarizations of these two thermal light beams are orthogonal in Fig. \ref{5}(a). There are eight different alternatives to trigger a two-particle coincidence count. However, the particles emitted by S$_1$ and S$_2$ are distinguishable. When the two particles are emitted by S$_1$ and S$_2$, respectively, there is no two-particle interference since these two different alternatives to trigger a two-photon coincidence count are distinguishable \cite{feynman,feynman-qpt}. It only contributes to the background coincidence count. Hence the peak in this condition is only half height of the one in Fig. \ref{4} \cite{liu-oe-2013}. The visibility of the dip in Fig. \ref{5}(a) is 22.22\%. When the polarizations of these two light beams are parallel, all the eight different alternatives are indistinguishable. The calculated second-order spatial coherence functions of two independent thermal fermion beams in a HOM interferometer are shown by the red circles in Fig. \ref{5}(b). The red curve is the fitting of the data by employing Eq. (\ref{HOM-F-3}). The size of both sources are assumed to be 0.59 mm in the fitting, which is calculated based on the curve in Fig. \ref{5}(a). The visibility of the fitted curve for fermions in Fig. \ref{5}(b) is 27.50\%, which is higher than the one of the dip in Fig. \ref{5}(a).

\section{Discussions}\label{discussions}

Comparing the calculated coincidence counts for \lq\lq fermionic photon\rq\rq  in Fig. \ref{3} of our experiments with the measured coincidence counts of thermal neutrons in Fig. 2 of Ref. \cite{iannuzzi}, it is easy to see the similarity between these two results. The spatial coherence functions of thermal fermions in Fig. \ref{4} of our experiments and Fig. 2 of Ref. \cite{jeltes} are also similar except the visibility is higher in our experiments. The reasons why the visibility of our measured second-order interference patterns is higher than the ones in Refs. \cite{iannuzzi,jeltes} are as follows. The response time of our detection system ($\sim$ 0.45 ns) is much shorter than the coherence time of thermal state (296 ns). The second reason is that the degeneracy factor of thermal state in our experiment is much larger than 1. If true thermal light, instead of pseudothermal light, is employed in our experiments, the visibility will be low \cite{zhai} as the ones in Refs. \cite{iannuzzi,jeltes}. Both temporal and spatial two-particle antibunching is obeserved for thermal fermions in a HBT interferometer, which means $g^{(2)}(0)$ is less than 1 for thermal fermions. Based on the nonclassical criterion that a state is nonclassical if $g^{(2)}(0)$ is in the range of $[0, 1)$  \cite{loudon}, thermal fermion state is a nonclassical state. There is one more thing worthy of noticing, the criterion of nonclassical state based on the value of $g^{(2)}(0)$ is only valid for a particle beam in a HBT interferometer. $g^{(2)}(0)$ of classical states in other interferometers can be less than 1. We have measured $g^{(2)}(0)=0.76\pm0.04$ by superposing thermal and laser light beams in a HOM interferometer \cite{liu-epl,liu-josaa}.

There are also second-order interference experiments with massive particles in a HOM interferometer \cite{liu,lopes}. In Ref. \cite{lopes}, Lopes \textit{et. al.} employed entangled $^4$He atom pairs to measure two-particle coincidence counts, which is a direct generalization of HOM dip \cite{hom} of photons to atoms. In Ref. \cite{liu}, two electron beams are incident to a HOM interferometer and electron current noise suppression is observed by measuring one of these two output signals. Even though they did not measure the correlation of two output signals, the noise level of one beam can be employed to predict the property of particle beam based on Mandel's Q factor \cite{mandel-book}. In the language of two-particle interference, the noise suppression in Liu \textit{et. al.}'s experiment is due to quantum interference of electrons in a HOM interferometer \cite{liu}.

\begin{figure}[htb]
\centering
\includegraphics[width=70mm]{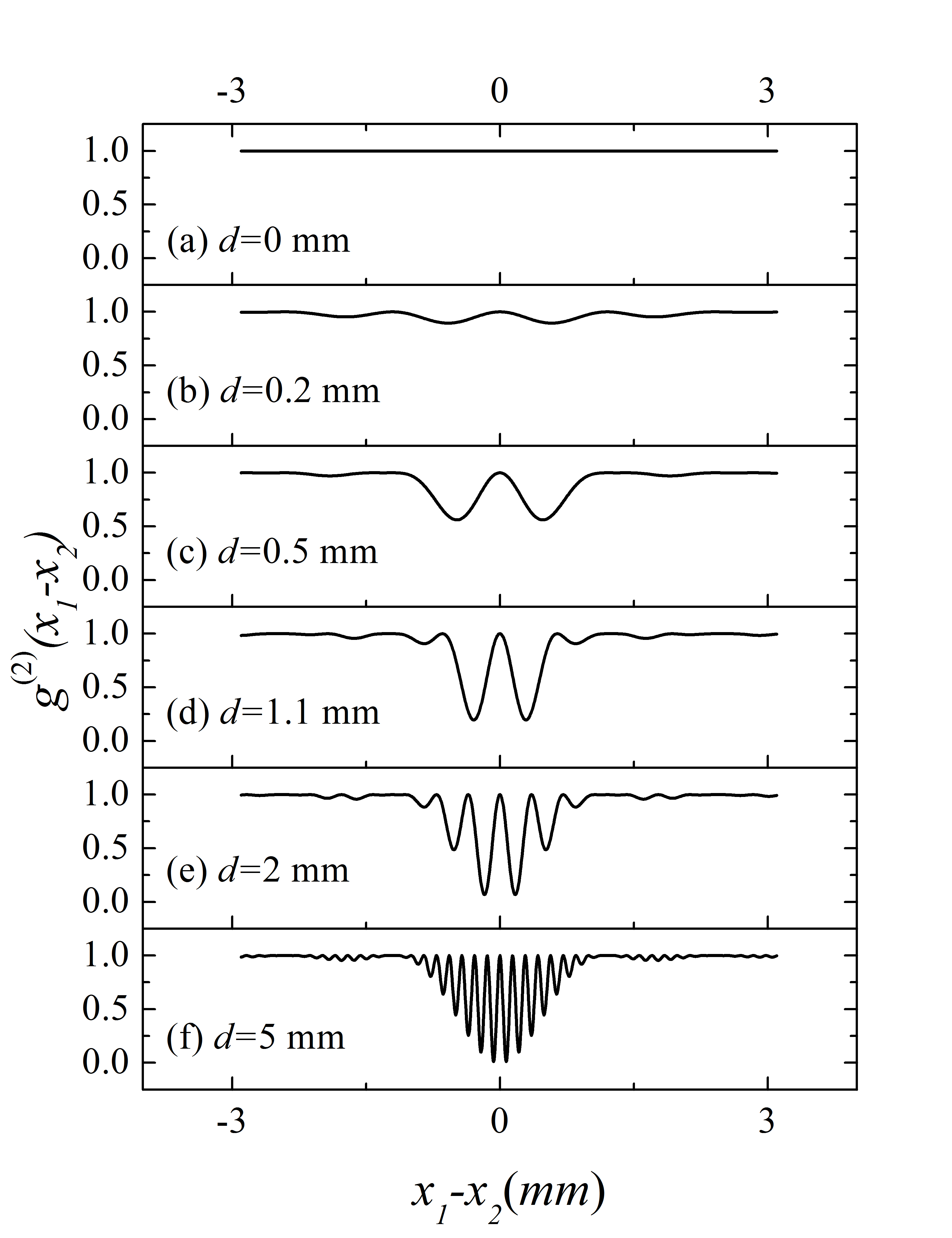}%
\caption{Simulated second-order interference patterns of two independent thermal fermion beams in a HOM interferometer based on Eq. (\ref{HOM-F-3}). $g^{(2)}(x_1-x_2)$ is the normalized second-order coherence function. $x_1-x_2$ is the relative distance between the transverse coordinates of D$_1$ and D$_2$, respectively. The simulation parameters are chosen to be the same as the ones in our experiments. $l$: 0.59 mm. $\lambda$: 780 nm. $z$: 910 mm. $d$: the varying parameter. }\label{6}
\end{figure}

In order to get a better understanding about the second-order interference of two independent thermal fermion beams in a HOM interferometer, we further analyze the experiments in Feynman's path integral theory.  When the two thermal fermion beams are different types of fermions or fermions with different spins that the detection system can distinguish, the measured second-order coherence function of two thermal fermion beams in a HOM interferometer is shown by the red curve in Fig. \ref{5}(a). The measured second-order coherence function is not a direct sum of two HBT dips. The reason why the visibility of the dip in Fig. \ref{5}(a) is less than the one of single HBT dip in Fig. \ref{4} is the coincidence count of two particles emitted by two sources, respectively, only contributes to the background coincidence counts.

When the particles emitted by these two sources are indistinguishable, all the eight different alternatives are indistinguishable. The second-order interference patterns are shown in Figs. \ref{5}(b) and \ref{6}(a)-(f). All the parameters in Fig. \ref{6}(d) are the same as the ones in our experiments and the only varying parameter in the simulation is the distance between these two sources. The size of these two sources both equal 0.59 mm. The distances between the source and detection planes are all 910 mm. The central wavelength of photon is 780 nm. When these two sources are in the symmetrical positions, no second-order interference pattern can be observed as shown in Fig. \ref{6}(a). Based on the results in Fig. \ref{6}, we can concluded that the two input electron beams in Liu \textit{et. al.}'s experiments \cite{liu} were not symmetrical, otherwise there will be no noise suppression. The observed suppression in electron current noise in Ref. \cite{liu} is a direct result of two-electron antibunching in a HOM interferometer. The smaller the value of $g^{(2)}(x_1-x_2)$ is, the larger the electron current noise suppression will be. $g^{(2)}(x_1-x_2)$ always equals 1 when these two sources are symmetrical, which means no two-particle antibunching and no electron current noise suppression can be observed. When the value of $d$ increases, larger two-particle antibunching can be obtained as shown in Fig. \ref{6}. For instance, $g^{(2)}(x_1-x_2)$ can equals 0.25 when $d$ equals 5 mm as shown in Fig. \ref{6}(e). The distance between these two sources in their experiments should not be too small since the noise level is about half of the one when only one electron beam is in the interferometer \cite{liu}.

One might argue that all fermions are massive, how can the interference patterns with massless \lq\lq fermionic photons \rq\rq be applied to real fermions. In Sect. \ref{theory}, we have proved that all types of thermal fermion beams satisfy Eq. (\ref{HBT-f-3}) when they are in a HBT or HOM interferometer. In Feynman's path integral theory, all the fermions obey the same superposition principle. The only difference between massive fermions and massless \lq \lq fermionic photon\rq \rq is they have different forms of propagators. One can calculate the second-order interference pattern by employing proper Feynman's propagator. The second-order interference pattern of massive fermions will be similar as the one with \lq \lq fermionic photon\rq \rq.

Just as the conclusion at the end of T\"{o}ppel \textit{et. al.}'s paper, it is possible to generalize some of the above results to the third- and higher-order interference of fermions. To the best of our knowledge, there is no third- or high-order interference experiment with fermions. However, there had been third- and higher-order interference experiments with photons \cite{zhou,pan}. It is an interesting topic to predict the third- and higher-order interference of fermions based on the results of photons.

\section{Conclusions}\label{conclusion}

In conclusion, we have proved that the second-order coherence function of non-identical thermal fermions in a HBT (HOM) interferometer can be represented by the coherence functions of bosons and classical particles. The calculated second-order interference patterns of \lq\lq fermionic photon\rq\rq are similar as the ones of massive fermions, which are also consistent with the theoretical predictions. The results in this paper are helpful understand the second-order coherence of fermions. The employed method offers a new way to experimentally study the coherence properties of fermions with photons, which is much more convenient than with massive fermions in contemporary technology.

\section*{Acknowledgement}

This project is supported by National Science Foundation of China (No.11404255), Doctoral Fund of Ministry of Education of China (No.20130201120013), the 111 Project of China (No.B14040) and the Fundamental Research Funds for the Central Universities.


\begin{thebibliography}{99}

\bibitem{note-anyons} There is another type of particle besides boson and fermion, named anyon, which has fractional spin. However, anyons can only exist in one or two dimension case. The particles consisting of our three-dimenional universe are either bosons or fermions. For instance, see J. M. Leinaas and J. Myrheim, IL Nuovo Cimento B \textbf{37}, 1 (1977).

\bibitem{dirac} P. A. M. Dirac, \textit{The Principle of Quantum Mechanics (4th ed.)} (Science Press, Beijing, 2008).

\bibitem{mandel-book} L. Mandel and E. Wolf, \textit{Optical Coherence and Quantum Optics} (Cambridge University Press, New York, 1995).

\bibitem{nielsen} M. Nielsen and I. Chuang, \textit{Quantum Information and Quantum Computation} (Cambridge University Press, New York, 2010).

\bibitem{aspect} P. Grangier, G. Roger, and A. Aspect, Europhys. Lett. \textbf{1}, 173 (1986).

\bibitem{hbt} R. Hanbury Brown and R. Q. Twiss, Nature \textbf{177}, 27 (1956).

\bibitem{bachor} Hans-A. Bachor and T. C. Ralph, \textit{A Guide to Experiments in Quantum Optics} (Wiley-VCH Verlag GmbH \& Co. KGaA, Weinheim, 2004).

\bibitem{bec-1} E. A. Cornell and C. E. Wieman, Rev. Mod. Phys. \textbf{74}, 875 (2002).

\bibitem{bec-2} W. Ketterle, Rev. Mod. Phys. \textbf{74}, 1131 (2002).

\bibitem{tonomura} A. Tonomura, J. Endo, T. Matsuda, T. Kawasaki, and H. Ezawa, Am. J. Phys. \textbf{57}, 117 (1989).

\bibitem{andrews} M. R. Andrews, C. G. Townsend, H.-J. Miesner, D. S. Durfee, D. M. Kurn, and W. Ketterle, Science \textbf{275}, 637 (1997).

\bibitem{mandel-laser} G. Magyar and L. Mandel, Nature \textbf{198}, 255 (1963).

\bibitem{schellekens} M. Shellekens, R. Hoppeler, A. Perrin, J. Viana Gomes, D. Boiron, A. Aspect, and C. I. Westbrook, Science \textbf{310}, 648 (2005).

\bibitem{glauber} R. J. Glauber, Phys. Rev. \textbf{131}, 2766 (1963).

\bibitem{henny} M. Henny, S. Oberholzer, C. Strunk, T. Heinze, K. Ensslin, M. Holland, and C. Schonenberger, Science \textbf{284}, 296 (1999).

\bibitem{oliver} W. D. Oliver, J. Kim, R. C. Liu, and Y. Yamamoto, Science \textbf{284}, 299 (1999).

\bibitem{liu} R. C. Liu, B. Odom, Y. Yamamoto, and S. Tarucha, Nature \textbf{391}, 263 (1998).

\bibitem{rom} T. Rom, Th. Best, D. van Oosten, U. Schneider, S. F\"{o}lling, B. Paredes, and I. Bloch, Nature \textbf{444}, 733 (2006).

\bibitem{iannuzzi} M. Iannuzzi, A. Orecchini, F. Sacchetti, P. Facchi, S. Pascazio, Phys. Rev. Lett. \textbf{96}, 080402 (2006).

\bibitem{jeltes} T. Jeltes, J. M. McNamara, W. Hogervorst, W. Vassen, V. Krachmalnicoff, M. Schellekens, A. Perrin, H. Chang, D. Boiron, A. Aspect, C. I. Westbrook, Nature \textbf{445}, 402 (2007).

\bibitem{schmidt} T. L. Schmidt, A. Komnik, and A. O. Gogolin, Phys. Rev. Lett. \textbf{98}, 056603 (2007).

\bibitem{neder} I. Neder, N. Ofek, Y. Chung, M. Heiblum, D. Mahalu, V. Umansky, Nature \textbf{448}, 333 (2007).

\bibitem{bocquillon} E. Bocquillon, V. Freulon, J.-M Berroir, P. Degiovanni, B. Pla\c{c}ais, A. Cavanna, Y. Jin, and G. F\`{e}ve, Science \textbf{339}, 1054 (2013).

\bibitem{zeilinger} A. Zeilinger, Phys. Scr. \textbf{T76}, 203 (1998).

\bibitem{omar} Y. Omar, N. Paunkovi\'{c}, L. Sheridan, S. Bose, Phys. Rev. A \textbf{74}, 042304 (2006).

\bibitem{peruzzo} A. Peruzzo, M. Lobino, and J. C. F. Matthews \textit{et.al.}, Science
\textbf{329}, 1500 (2010).

\bibitem{sansoni} L. Sansoni, F. Sciarrino, G. Vallone, P. Mataloni, A. Crespi, R. Ramponi, and R. Osellame, Phys. Rev. Lett. \textbf{108}, 010502 (2012).

\bibitem{matthews} J. C. F. Matthews, K. Poulios, J. D. A. Meinecke, A. Politi, A. Peruzzo, N. Ismail, K. W\"{o}rhoo, M. G. Thompson, and J. L. O'Brien, Sci. Rep. \textbf{3}, 1539 (2013).

\bibitem{tichy} M. C. Tichy, J. Phy. B: At. Mol. Opt. Phys. \textbf{47}, 103001 (2014).

\bibitem{crespi} A. Crespi, L. Sansoni, G. Della Valle, A. Ciamei, R. Ramponi, F. Sciarrino, P. Mataloni, S. Longhi, and R. Osellame, Phys. Rev. Lett. \textbf{114}, 090201 (2015).

\bibitem{hubel} H. H\"{u}bel, D. R. Hamel, A. Fedrizzi, S. Ramelow, K. J. Resch, and T. Jennewein, Nature \textbf{446}, 601 (2010).

\bibitem{toppel} F. T\"{o}ppel and A. Aiello, Phys. Rev. A \textbf{88}, 012130 (2013).

\bibitem{feynman} R. Feynman, R. B. Leighton, and M. L. Sands, \textit{The Feynman's Lectures on Physics vol. 3} (Pearson Education Asia Inc. and Beijing World Publishing Crop., Beijing, 2004).

\bibitem{feynman-qpt} R. P. Feynman and A. R. Hibbs, \textit{Quantum Mechanics and Path Integral} (McGraw-Hill, Inc., New York, 1965).

\bibitem{liu-oc-2015} J. B. Liu, Y. Zhou, H. B. Zheng, H. Chen, F. L. Li, and Z. Xu, Opt. Commun. \textbf{354}, 79 (2015).

\bibitem{liu-cpb} J. B. Liu, D. Wei, H. Chen, Y. Zhou, H. B. Zheng. H. Gao, F. L. Li, and Z. Xu, Chin. Phys. B \textbf{25}, 034203 (2016).

\bibitem{hbt-1} R. H. Brown and R. Q. Twiss, Nature \textbf{178}, 1046 (1956).

\bibitem{shih-book} Y. H. Shih, \textit{An Introduction to Quantum Optics} (Taylor and Francis Group, LLC, FL, 2011)

\bibitem{note-f} Here we employ the concept \lq\lq fermionic photon\rq\rq to mean a special type of fermion, whose properties, such as massless, frequency, wavelength, are the same as photon except they obey different statistics.

\bibitem{shih} Y. H. Shih and C. O. Alley, Phys. Rev. Lett. \textbf{61}, 2921 (1988).

\bibitem{hom} C. K. Hong, Z. Y. Ou, and L. Mandel, Phys. Rev. Lett. \textbf{59}, 2044 (1987).

\bibitem{santori} C. Santori, D. Fattal, J. Vu\u{c}kovi\'{c}, G. S. Solomon, and Y. Yamamoto, Nature \textbf{419}, 594 (2002).

\bibitem{liu-oe-2013} J. B. Liu, Y. Zhou, W. T. Wang, R. F. Liu, K. He, F. L. Li, and Z. Xu, Opt. Express \textbf{21}, 19209 (2013).

\bibitem{loudon} R. Loudon, \textit{The Quantum Theory of Light (3rd ed.)} (Oxford University Press, New York, 2000).

\bibitem{martienssen} W. Martienssen and E. Spiller, Am. J. Phys. \textbf{32}, 919 (1964).

\bibitem{shapiro} J. H. Shapiro and R. W. Boyd, Quantum Inf Process \textbf{11}, 949 (2012).

\bibitem{note} The measured coherence time in Fig. 1 (296 ns) is not the coherence time we employed in our spatial experiments. In spaital one, we changed the rotation speed of the ground glass so that the coherence time of the pseduthermal light is about 90 $\mu$s. Hence the coincidence time window is much less than the coherence time.

\bibitem{zhai} Y. H. Zhai, X. H. Chen, D. Zhang, and L. A. Wu, Phys. Rev. A \textbf{72}, 043805 (2005).

\bibitem{liu-epl} J. B. Liu, Y. Zhou, F. L. Li, and Z. Xu, Europhys. Lett. \textbf{105}, 64007 (2014).

\bibitem{liu-josaa} J. B. Liu, Y. Zhou, F. L. Li, and Z. Xu, J. Opt. Soc. Am. A \textbf{31}, 1481 (2014).

\bibitem{lopes} R. Lopes, A. Imanaliev, A. Aspect, M. Cheneau, D. Boiron, and C. I. Westbrook, Nature \textbf{520} 66 (2015).

\bibitem{zhou} Y. Zhou, J. Simon, J. B. Liu, and Y. H. Shih, Phys. Rev. A \textbf{81}, 043831 (2010).

\bibitem{pan} J. W. Pan, Z. B. Chen, C. Y. Lu, H. Weinfurter, A. Zeilinger, and M. \.{Z}ukowski, Rev. Mod. Phys. \textbf{84}, 777 (2012).

\end{thebibliography}
\end{document}